\documentclass[12pt, draftclsnofoot, onecolumn]{IEEEtran}
\usepackage{amsmath,amsfonts}
\usepackage{algorithmic}
\usepackage{algorithm}
\usepackage{array}
\usepackage[caption=false,font=normalsize,labelfont=sf,textfont=sf]{subfig}
\usepackage{textcomp}
\usepackage{stfloats}
\usepackage{url}
\usepackage{verbatim}
\usepackage{graphicx}
\usepackage{cite}
\usepackage{xcolor}
\graphicspath{{./figures/}}

\begin{document}

\title{Wireless Interconnection Network (WINE) for Post-Exascale High-Performance Computing}

\author{
    Hong Ki Kim, Yong Hun Jang, Hee Soo Kim, \\Won Young Kang, Young-Chai Ko, and Sang Hyun Lee
    \thanks{This work was supported in part by the National Research Foundation of Korea (NRF) Grant funded by the Korea Government (MSIT) under Grant 2022R1A5A1027646, and in part by the Institute of Information and Communications Technology Planning and Evaluation (IITP) Grant funded by the MSIT under Grant 2021-0-00260.}
    \thanks{All authors are with the School of Electrical Engineering, Korea University, Seoul 02841, Korea (e-mail:\{istackcheese, disclose, huolangheesoo, dogs0667, koyc, sanghyunlee\}@korea.ac.kr).}
}

\maketitle

\begin{abstract}
Interconnection networks, or `interconnects,' play a crucial role in administering the communication among computing units of high-performance computing (HPC) systems. Efficient provisioning of interconnects minimizes the processing delay wherein computing units await information sharing between each other, thereby enhancing the overall computation efficiency. Ideally, interconnects are designed with topologies tailored to match specific workflows, requiring diverse structures for different applications. However, since modifying their structures mid-operation renders impractical, indirect communication incurs across distant units. In managing numerous long-routed data deliveries, heavy burdens on the network side may lead to the under-utilization of computing resources. In view of state-of-the-art HPC paradigms that solicit dense interconnections for diverse computation-hungry applications, this article presents a versatile wireless interconnecting framework, coined as Wireless Interconnection NEtwork (WINE). The framework exploits cutting-edge wireless technologies that promote workload adaptability and scalability of modern interconnects. Design and implementation of wirelessly reliable links are strategized under network-oriented scrutiny of HPC architectures. A virtual HPC platform is developed to assess WINE's feasibilities, verifying its practicality for integration into modern HPC infrastructures.
\end{abstract}

\clearpage

\section{Introduction}
Since the advent of supercomputers in the 1960s, advances in integrated circuits, computer engineering, and communication networks have led to significant evolution in high-performance computing (HPC). Surges of scientific data empower HPC as a major enabler of research breakthroughs to this day \cite{MdcConvergence}. State-of-the-art information technologies and their sophisticated applications call for progressively enhanced processing capability. To such demands, brand-new HPC architectures with extended compute nodes and improved networking functionalities continuously emerge. With ongoing competitions among the top-tier HPC systems, TOP500 project \cite{top500} records regular ranking report of the 500 most advanced computing technologies. As of 2024, Oak Ridge National Laboratory has most recently topped the list with benchmark performance over 1.206 exa-floating-point operations per-second (EFLOPS), paving ways towards imminent post-exascale race.

HPC architecture has experienced an exponential growth in the number of processing cores, evolving into massively interconnected clusters of computing resources. Parallel processing techniques are leveraged to decompose intense HPC tasks into multiple subtasks, distributing them throughout the system to expedite the job completion. This expands the focus of the computing performance from single-core processing power to intricate collaboration among individual processing units. Sophisticated coordination of interconnected computing units surpasses beyond scaling limitations of the semiconductor technology, characterized by the power wall and the memory wall \cite{walls}.
Efficient coordination of computing resources gives rise to versatile link management across large-scale HPC systems. Consequently, future advances in HPC capabilities pose network-oriented challenges of skillfully provisioning communication resources.

Interconnection network, or simply `interconnect,' is a critical component that unites separate computing units into a single cooperative cluster. Parallel processing over interconnects often requires data sharing between units without direct routes. Frequent interactions across multiple hops can overburden the interconnect, hindering efficient utilization of computing resources. To alleviate such network loads, physical structures of interconnects are designed to align with data stream patterns of target applications. Widely adopted interconnect topologies like fat-tree and torus \cite{9248644} are generally well-suited for a variety of applications in this regard.

A single interconnect structure, however, faces limitations in optimizing for diverse use cases. As new applications emerge and interconnects scale, the disparity between physical topologies and data sharing patterns widens. This structural mismatch leads to network overheads and data traffic congestion in large-scale clusters, becoming a major performance bottleneck in HPC operations \cite{tail_scale, Fujiwara2015}. Hence, the focus of next-generation HPC technology centers towards adaptive and versatile interconnect link maneuvering solutions.

Noteworthy and comparable to modern HPC systems are mega data centers (MDCs) \cite{MdcConvergence}. These interconnected server systems run assorted data-driven cloud services, going beyond traditional file-storage features. MDC clouds supervise extensive processing and data streams across system components, with noticeable similarities to HPC operations. Albeit originating from distinct purposes, HPCs and MDCs become increasingly similar in architectures and workflows \cite{MdcConvergence}, due to recent technology business trends of pervasive and computation-intensive services. Multifaceted coverage of MDC service catalogues suggests that this convergence of HPCs and MDCs will lead to the diversity and the scale of future HPC applications. Sustainable HPC administration requests adept large-scale interconnect that supports ever-evolving workflows.

This article introduces a promising interconnect framework called Wireless Interconnection NEtwork (WINE). WINE enhances the connectivity among computing units with wireless links, offering dynamic interconnect adaptivity. For practical deployment in real-world HPC systems, these wireless links necessarily provide high throughput and reliability. Therefore, the framework is designed with a comprehensive scrutiny of HPC architectures and their interconnect utility requirements.
A virtual test platform, `WINE-cellar,' is developed to prototype various design cases and assess their feasibility. The validation of key performances, from wireless signal strengths to benchmark processing speeds, confirms WINE's suitability for the on-site deployment.

The rest of this article proceeds as:
The first section overviews network requirements for scalable HPC systems.
Next, interconnection principles to meet these requirements are examined.
Subsequent section investigates candidate wireless technologies and strategies for deployment in HPC interconnects. 
Furthermore, link reliability and performance utilities are analyzed through extensive WINE-cellar prototyping verification.
Finally, the article concludes with discussions on ongoing research challenges.

\section{Interconnects for scalable HPCs}
\label{sec:hpc_system}

\begin{figure}
    \centering
    \includegraphics[width=\linewidth]{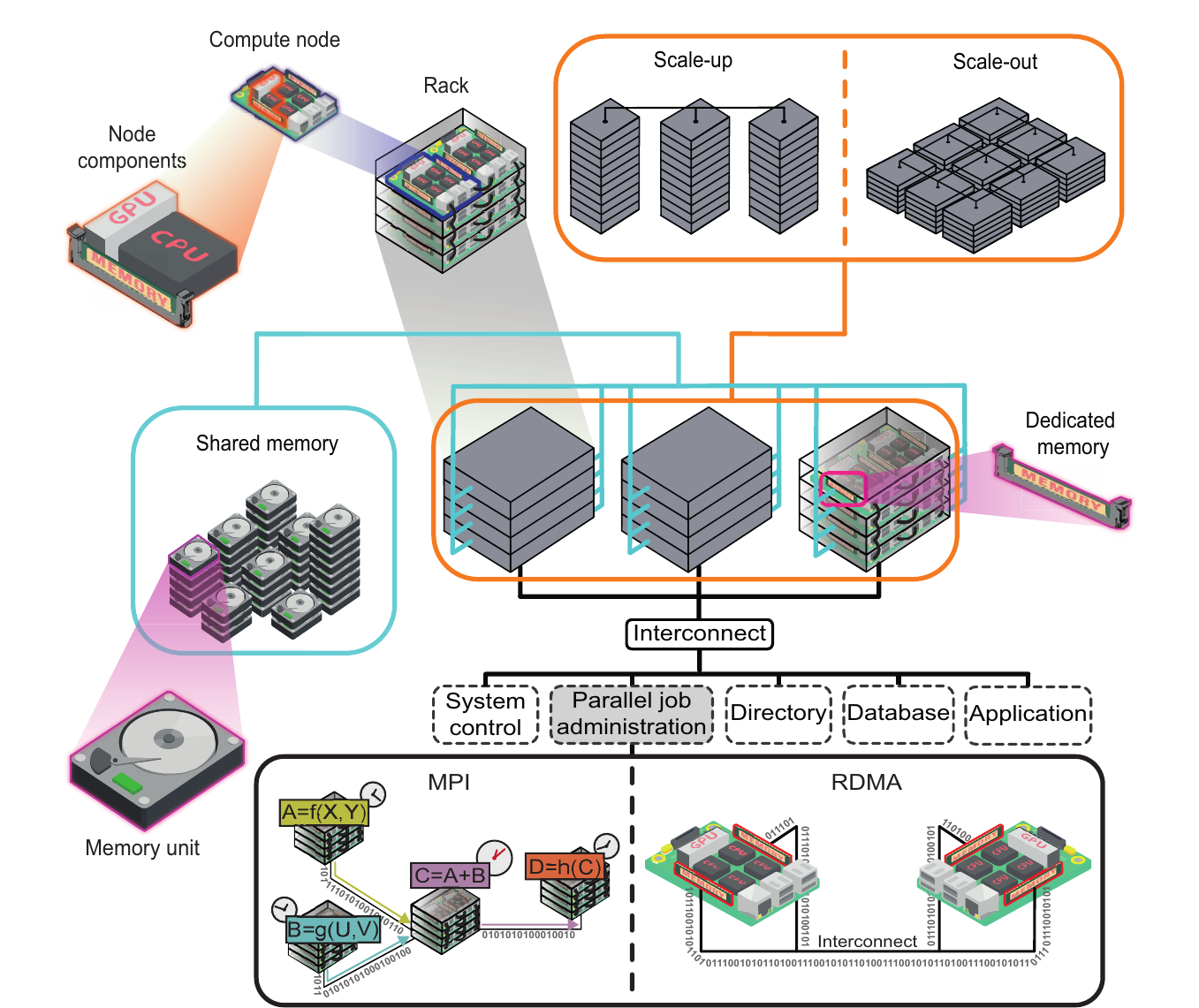}
    \caption{Modern HPC architecture, administration, and scaling.}
    \label{fig:hpc_architecture}
\end{figure}

Semiconductor technology advances have led to exponential growth in central processing unit (CPU) performance. This has propelled the evolution of HPC architectures, which heavily rely on parallel computing principles and massive CPU clusters. The design philosophies behind such architectures emphasize sophisticated processor structuring and interconnection strategies. Figure \ref{fig:hpc_architecture} depicts a modern HPC architecture, highlighting complex interconnections between components. The fundamental computing units of the system are compute nodes, each equipped with essential computing components. These components include CPUs, dedicated memories, and accelerators such as graphics processing units. Compute nodes are densely packaged into racks for efficient use of space, streamlined management, and effective cooling. High-speed interconnect joins the racks to the maintenance and interfacing units, such as the shared memory pool and the directory permission manager.

To incorporate more computing resources into the system, additional compute nodes can {\em scale-up} within each rack.
However, continued scale-up of processor throughput faces a number of engineering challenges. In particular, the memory wall \cite{walls} envisions a processing capability mismatch between CPU and memory components. This can cause CPUs to wait for memory operations to complete, resulting in the under-utilization of computing resources. On the other hand, increased power leakage from densified chip integration is referred to as the power wall \cite{walls}. The limited power delivery to processors bounds achievable computing performances.
Excessive scale-up can lead to overheating, thereby prompting the {\em scale-out} approach that accommodates additional racks. These scaling strategies have led to the interconnection of a vast number of compute nodes.

Massive HPC tasks are divided into smaller pieces and distributed across compute nodes. These distributed jobs are coordinated with message passing interface (MPI) \cite{MdcConvergence}, which employs periodic synchronization to maintain the coherence of memory and process. Long-routed communication between remote compute nodes can prolong the delay of MPI synchronization steps. Remote direct memory access (RDMA) is adopted for efficient memory consistency management, enabling direct access to the memory of each other among the nodes  \cite{9248644}.
Maintaining the inter-node process coherence is facilitated by delicate provisioning of interconnect resources. Hence, the link coordination in dense HPC networks engages with advanced interconnect technologies.

\begin{figure}
    \centering
    \includegraphics[width=0.95\linewidth]{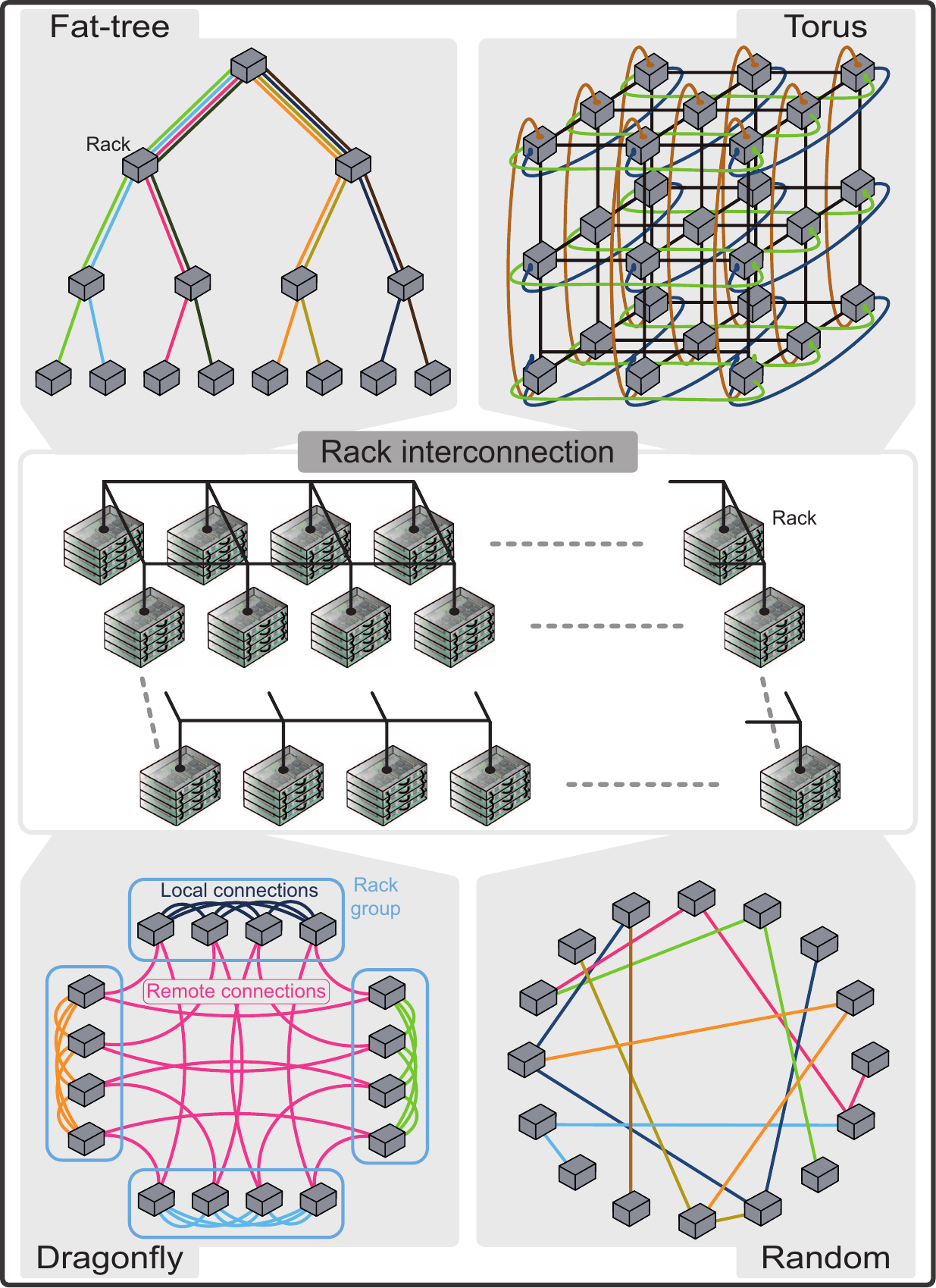}
    \caption{HPC interconnection principles.}
    \label{fig:topologies}
\end{figure}

Networking functionalities of advanced interconnects pursue two-fold objectives. First, they ensure reliable and fast data exchanges among compute nodes. On-site interconnects, such as InfiniBand and Gigabit Ethernet \cite{top500}, use high-end network interfaces to achieve data rates of up to hundreds of Gbps. Second, interconnects organize large groups of compute nodes into efficient structures for the job distribution. Leading HPC systems employ carefully designed interconnect topologies to meet these objectives. The topologies address various factors such as average routing distances, potential network congestion points, and link cable costs.

Figure \ref{fig:topologies} illustrates noteworthy interconnect topologies.
An $r$-radix fat-tree organizes racks into a tree topology, where each parent rack connects to $r$ child racks. Since the tree structure imposes heavy network loads on links near the root, link bandwidths are allocated in proportion to tree levels, resulting in {\em fatter} branches at upper hierarchies.
Routing paths between some rack pairs are considerably longer than others, thereby increasing synchronization delays. To mitigate this, programming that organizes processes into local groups helps minimizing the communication through long-distance routes.
On the other hand, torus achieves uniform routing path lengths between all racks. A $k$-ary $n$-cube torus arranges racks into an $n$-dimensional lattice with $k$ racks along each edge. Each rack connects to the racks on its opposite edges, resulting in $2k$ neighboring connections for every rack. This dense connectivity provides short latencies and numerous routing options for all source-destination pairs, at the cost of increased network infrastructure expenses and wiring complexities.

Dragonfly \cite{9248644} uses a combination of local and remote cables to modularize racks into tightly-knit groups. Remote connections employ optical fibers to provide high-bandwidth links, compensating for latencies over long distances. This hierarchical arrangement supports structured load balancing and offers short inter-rack routes at expense of additional layers of parallel programming complexity.
Random topology \cite{random_topology_hpc} links arbitrarily selected rack pairs, creating shortcuts that bypass long routing paths. While the physical implementation of random connections faces challenges due to complicated cabling, empirical studies show that they significantly reduce latencies in large-scale networks \cite{random_topology_hpc}.
This motivates the development of network science-based WINE link coordination principles to reinforce the connectivity and scale the HPC performance by providing short routes between remote racks over scaled-out interconnects \cite{barabasiTemporal}.

\section{Wireless Interconnection Network (WINE)}
\label{sec:wireless}
This section presents the WINE framework and practical strategies to incorporate advanced wireless technologies.

\subsection{Why wireless?}
\label{sec:why_wireless}
Currently deployed HPC interconnects feature dense connectivity among computing units to achieve EFLOPS-level performance.
To evaluate the HPC system computing capability, standardized parallel processing benchmarks, such as NAS parallel benchmarks \cite{npb}, are utilized. Such representative tasks and other diverse real-world applications that emerge to this day \cite{MdcConvergence} require extensive cooperation among interconnected computing units. This entails the global cooperation among non-neighboring units through remote routing between units separated by a large number of hops.
In scaled-out systems, such latencies over long routes become orders of magnitude greater than data processing times.
Prolonged tail latencies \cite{tail_scale}, characterized by the critical path of the network, can bottleneck significant portions of compute nodes from consecutive processing. Thus, effective use of the computing resources in large-scale interconnects solicits the ability to dynamically form direct links between arbitrary units.

Advanced high-frequency wireless technologies enable the realization of wireless links with data rates reaching hundreds of Gbps \cite{6G}. These high-throughput capabilities match those of current top-tier interconnect links, ensuring that wireless links do not pose any bottleneck when deployed alongside traditional wired links. Practical challenges in utilizing such high-frequency spectra are severe atmospheric attenuation and molecular absorption in outdoor mobile environments. Fortunately, controlled indoor environment of HPC facilities lifts major concerns of the propagation loss. 
Furthermore, beamforming techniques of high-frequency radio spectra \cite{mmWave, Rikkinen2020} and high spatial resolution of optical signals \cite{optics_for_DC} can significantly reduce the interference in dense wireless interconnection environments. By focusing the signal energy in the intended direction, narrow-beam characteristics are exploited to mitigate beam collisions. With the prior information of rack positions, accurate beam training \cite{mmWave} can configure optimal steering directions towards receiving ends. While the training procedure is periodically conducted in typical wireless networks with mobile agents, one-time pre-training suffices for HPC environments with stationary infrastructures only. The job administration in WINE leverages the pre-trained beam steering angles for swift and accurate inter-rack data sharing with the minimal interference.

Augmenting interconnects with high-speed wireless links, direct routes can be created between remote parts of the scaled-out system. This capability significantly enhances the coordination between computing resources, reducing the job administration latency for expedited processing. Furthermore, dynamic tailoring of the interconnect to diverse applications is facilitated via flexible scheduling of the job assignment to compute nodes, a process termed job mapping \cite{Fujiwara2015}. Fundamental HPC operations like double-precision floating-point calculations and large-size message exchanges involve different job mappings to optimize the computing resource utilization. The presented framework dynamically supports multiple topologies, accommodating various job mappings for efficient coordination of compute nodes.
Candidate frequency spectra for wireless links and strategies to integrate them into the HPC housing space are explored in the following subsection.

\subsection{WINE-making principle}
\label{sec:enabler}

\begin{figure}
    \centering
    \includegraphics[width=0.95\linewidth]{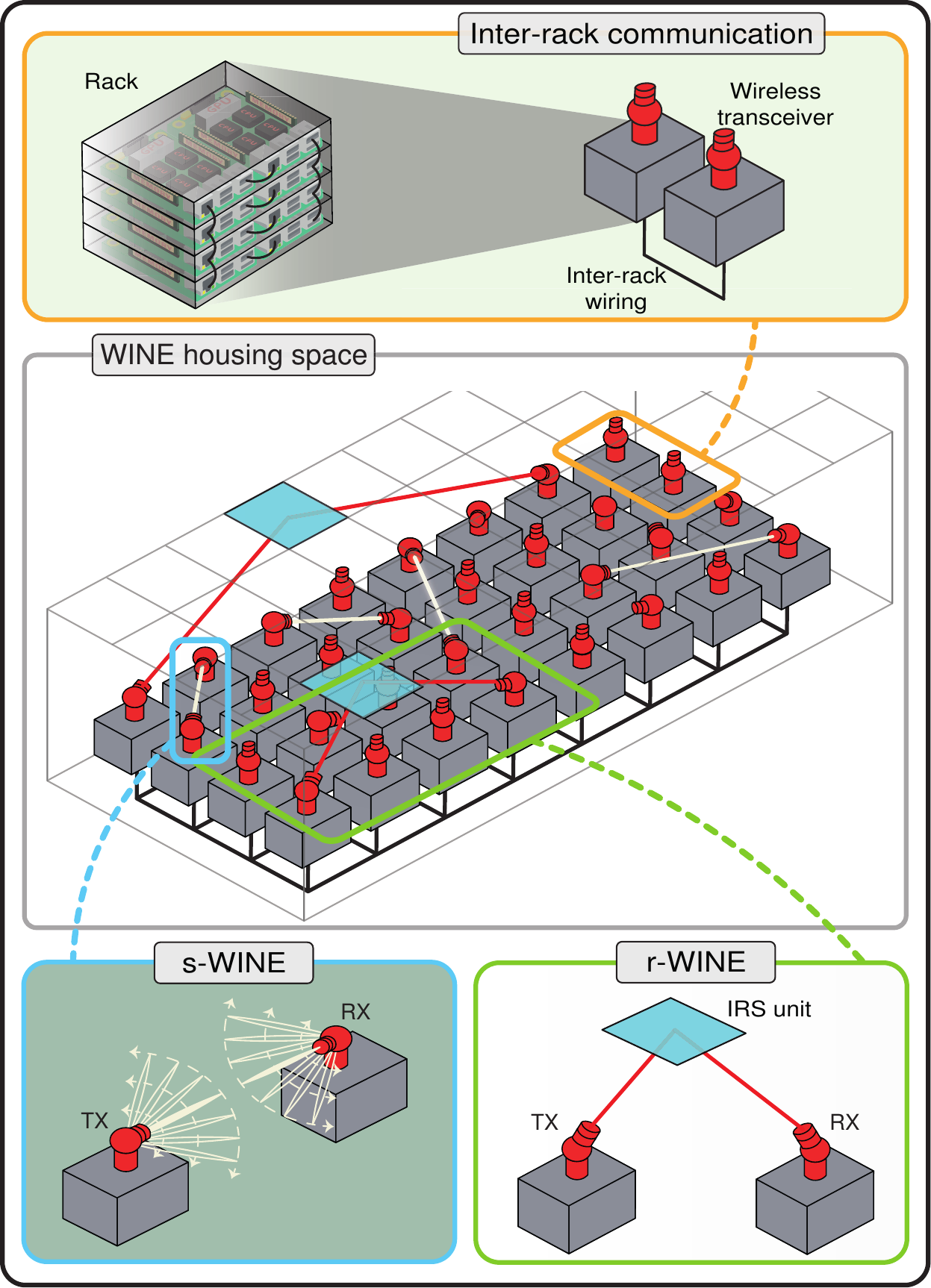}
    \caption{WINE-integrated HPC platform.}
    \label{fig:wine_schematics}
\end{figure}

The WINE framework can incorporate various broadband technologies to implement wireless links. Wireless links must seamlessly complement their wired counterparts to ensure sufficient reliability and data rates. Furthermore, dense wireless interconnections face challenges in minimizing the link interference. Meeting such demands with contemporary communication standards presents several research challenges.
For example, 3GPP Release 15 implements narrow-beam signals of several Gbps via millimeter-wave technologies \cite{mmWave}.
The adoption of 60GHz bands in IEEE 802.11ad for wireless MDC interconnects is a topic of ongoing research \cite{mmWave}. On the other hand, FSO technologies offer data rates comparable to those of HPC interconnects, and the utilization of optical links for the network reconfiguration are studied \cite{Fujiwara2015, optics_for_DC}. Terahertz-frequency bands have also gained the attention as promising candidate means for broadband transmission \cite{Rikkinen2020}. Studies on terahertz communication standards address several use cases of hyper-connected machine-type networks with key performance indicators, such as peak data rate, latency, and connection density \cite{6G}. Dual spectral properties between electromagnetism and photonics allow the expertise from both domains in realizing terahertz hardware technologies for the data transmission.

Figure \ref{fig:wine_schematics} illustrates a design case of WINE-integrated HPC system. Compute nodes, packaged into racks, are housed within an electromagnetically shielded chamber that prevents the interference with wireless signals of external sources. Inter-rack wiring provides the baseline connectivity that is exploited along with wireless links. Each rack is equipped with a wireless transceiver that operates in candidate spectrum ranges.
Potential implementation strategies for these transceivers include diverse antenna architectures, such as multi-antenna arrays and lens antennas that achieve high directionality \cite{6G}. Practical constraints in physical dimensions are ensured for the installation on the rack without causing significant signal obstruction.

The WINE framework employs two types of wireless links. First, direct wireless links, referred to as {\em straight} WINE (s-WINE), can establish line-of-sight connections between two transceivers. Notably, signals between some transceivers may be obstructed by dense infrastructure placement, or interfered by other signals. To enhance the degrees of freedom in wireless link configurations, WINE leverages intelligent reconfigurable surface (IRS) technologies \cite{IRS_tutorial}. An array of IRS units, consisting of computer-controlled metasurfaces, is installed on the ceiling of the housing structure. Each unit tweaks physical properties of the reflected signal to deliver the signal to the intended receiver.
Such wireless links that utilize the controlled reflective environment are referred to as {\em reflected} WINE (r-WINE) links. Careful placement of IRS units, transceivers, and racks can accommodate unobstructed non-interfering links, making efficient use of the limited space of the HPC facility.

A collective control for the s-WINE, r-WINE, and wired links facilitates the interconnect reconfiguration that surpasses beyond the restrictions of cabling geometry.
Direct links between distant rack pairs effectively reduce end-to-end latencies across large-scale interconnects, tailoring the network structure to the desired workload administration.
Also, diverse networking properties of traditional topologies can be exploited by the consolidation of their link types. The s-WINE and r-WINE links reinforce the data throughput of inter-rack connections that are taxed with heavy network loads. The link-specific bandwidth adjustment can alleviate network congestions, similarly to fat-trees. Regular neighboring rack connections of tori are naturally similar to the grid-like cabling layout, to which wireless links supplement direct connections between non-neighboring racks. Combinations of wired and wireless links attain dense connectivity of Dragonfly's local groups, and remote inter-group connections can be wirelessly configured. Thus, versatile interconnect management across various topologies ensures reliable networking among computing units.

\section{Feasible design case study}
\label{sec:feasibility}
This section verifies feasible design and deployment cases of WINE. 
In-depth performance analysis of the framework is presented from individual signal reliability to system-wide interconnect utilities.

\subsection{WINE link implementation}
\label{sec:link_budget}
The coexistence of wired and wireless links entails a balanced range of data rates for both link types. This warrants the reliability in data exchanges across compute nodes, encapsulating physical-layer details to focus on the data link coordination. To ensure that such physical utilities are established, a link budget analysis is conducted on wireless links implemented with the candidate frequency spectra discussed in Section \ref{sec:why_wireless}, within detailed HPC housing setups.
The analysis adopts spectral parameters of terahertz and optical frequencies for indoor wireless communication \cite{Rikkinen2020, 8360928} to simulate realistic propagation environments of HPC facilities. Currently deployed wired interconnects, such as InfiniBand, announce the single-lane throughput of 100Gbps in the latest product roadmaps \cite{top500}. Noting that wired interconnects often deploy multiple parallel cables for enhanced bisection bandwidths, the link budget analysis targets link capacities of hundreds of Gbps. This ensures consistent performance when wired and wireless link types are used together.

\begin{table*}
\caption{Link budget analysis for WINE links.}
\label{tab:link_budget}
\centering
\resizebox{0.95\textwidth}{!}{
\begin{tabular}{|c||m{7em}|m{7em}|m{7em}|m{7em}|}
\hline
Link type & THz (near) & THz (far) & FSO (near) & FSO (far) \\ \hline
Frequency (THz) & 0.30	& 0.30	& 193.55 & 193.55 \\ \hline
Bandwidth (GHz) & 30 & 30	& 50 & 50 \\ \hline
Distance (m) & 2 & 28.28 & 2 & 28.28  \\ \hline
Propagation loss (dB) & 88.01 & 111.02 & - & - \\ \hline
Geometric loss (dB) & - & - & 0.40 & 4.39 \\ \hline
Pointing loss (dB) & - & - & 5 & 5 \\ \hline
Atmospheric loss (dB) & 0.008 & 0.034 & 0 & 0 \\ \hline
Number of antenna	& 32x32 & 32x32 & 1 & 1 \\ \hline
% 4x4, 12.04 dB
TX power (dBm)	& 13 & 13 & 23.01 & 23.01 \\ \hline
TX gain (dBi) & 32.1 & 32.1 & - & - \\ \hline
RX gain (dBi) & 32.1 & 32.1 & - & - \\ \hline
Lens diameter (mm) & - & - & 24 & 24 \\ \hline
Beam divergence (mrad) & 123.92 & 123.92 & 0.28 & 0.28 \\ \hline
Noise figure (dB) & 12 & 12 & 15 & 15 \\ \hline
Total noise power (dBm) & 57.23 & 57.23 & 6.45 & 6.45 \\ \hline
SNR (dB) & 42.82 & 19.78 & 21.07 & 17.07  \\ \hline
Modulation & 16-QAM & 16-QAM & PAM & PAM\\ \hline
Achievable capacity (Gbps) & 426.73 & 197.62 & 350.51 & 284.96 \\ \hline
\end{tabular}}
\end{table*}

Table \ref{tab:link_budget} presents the wireless link budget analysis. Channel utilities of terahertz and FSO signals for both near and far inter-transceiver distances are analyzed under standard room temperature and indoor atmospheric conditions. A link between two transceivers of neighboring racks has the distance of $2m$. In a $20m\times20m\times5.5m$ machine room, the farthest diagonal link distance is $28.28m$.
To compensate for larger propagation loss from longer distances, wireless links directed to farther receiving ends utilize increased transmission powers. Industry-deployed form-factor HPC racks necessarily mount compact transceivers such as 32-by-32 terahertz multi-antenna arrays. The link budget analysis verifies that wireless links achieve 100Gbps-order data rates in the HPC housing environment. Thus, both link realization strategies over near and far interconnection distances sufficiently cover multi-lane bandwidths of currently deployed wired interconnects. This validates the suitability of WINE for the deployment within state-of-the-art HPC systems (see https://youtu.be/2z6r5ASJAlE).

\subsection{Housing dimensions for dense WINE links}
\label{sec:link_simul}

Dense wireless connections in compact HPC housing space are subject to the physical obstruction and the mutual interference. To configure wireless links beyond the infrastructure geometry, the r-WINE links exploit virtual line-of-sight propagation via IRS units. Given deterministic locations of transceivers and IRS units, accurate channel state information between two units can be easily obtained. Furthermore, lists of panels that provide strong r-WINE signals for individual source-destination transceiver pairs can be identified. Upon inter-rack data sharing requests, the lists consolidate a simple best-effort protocol to select a panel for the highest throughput while managing non-overlapping panel assignments.

\begin{figure}
    \centering
    \includegraphics[width=\linewidth]{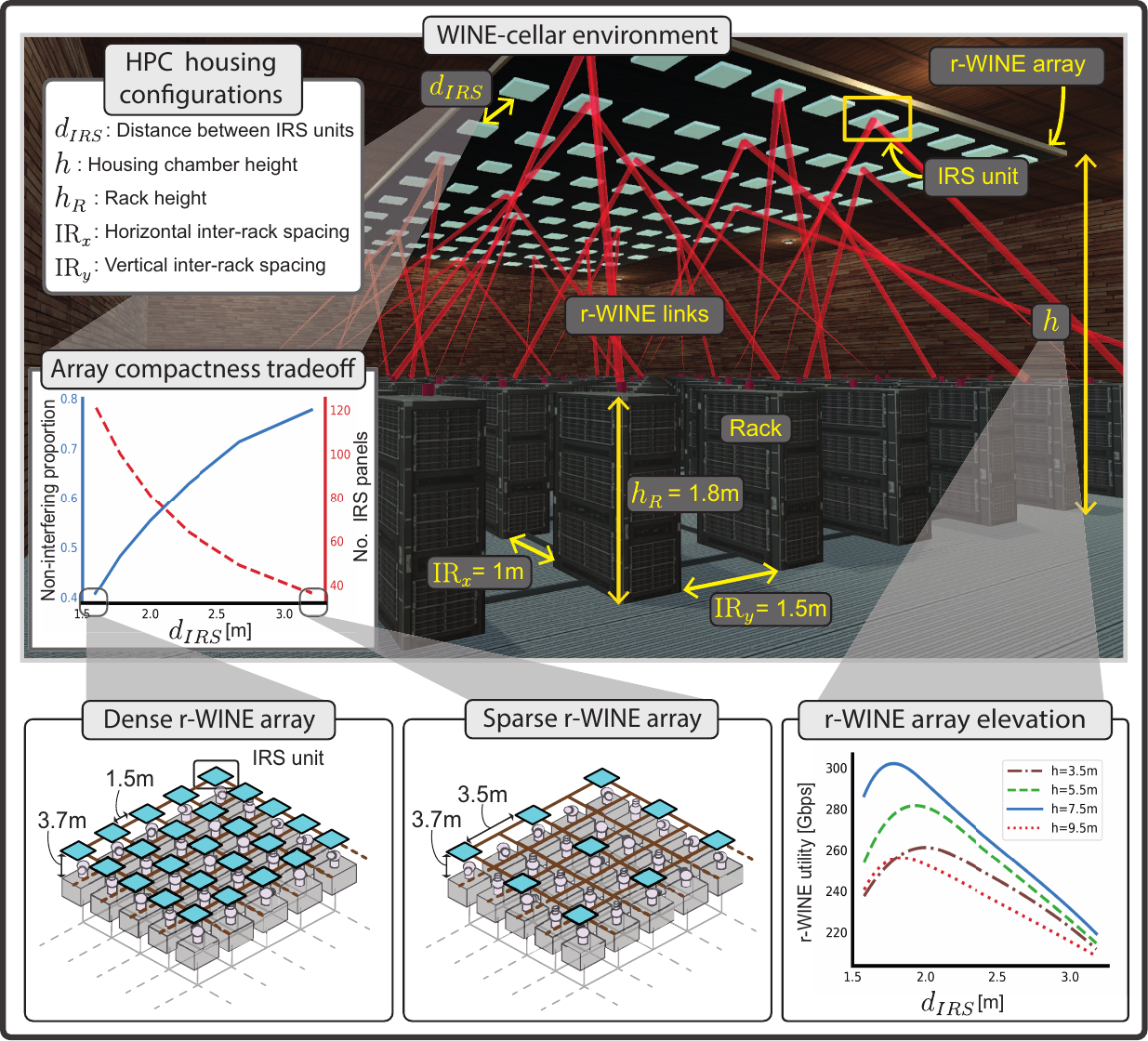}
    \caption{Digital twin HPC environment and WINE utilities.}
    \label{fig:rwine_config}
\end{figure}

For the development of qualified environments with dense wireless connections, various system configurations, such as housing dimensions, rack arrangement, and compactness of IRS units, intricately affect r-WINE link utilities. The digital twin platform, WINE-cellar, facilitates fine-tuning such design issues through virtual HPC system prototyping.
Figure \ref{fig:rwine_config} illustrates a WINE-cellar prototyping instance, its configurations, and the resulting r-WINE utilities.
The showcased virtual HPC prototype houses $10\times10$ rack layout in $20m\times20m$ machine room. These settings offer a compact and realistic environment, as recently reported TOP500 HPCs mostly consist of tens to hundreds of racks, with the current top-ranked system comprising 74 racks \cite{top500}.

Among HPC housing configurations, the regular spacing between IRS units, denoted by $d_{IRS}$, introduces a trade-off between the number of simultaneously deployable r-WINE links and effective interference management. This relationship is observed by adjusting $d_{IRS}$ from $1.5m$ to $3.5m$ along $18m\times18m$ flat-surfaced r-WINE array installed on the ceiling. Note that IRS units can be manufactured in sufficiently compact sizes \cite{IRS_tutorial}, allowing for the accommodation of 121 units under the densest r-WINE array setup, $d_{IRS}=1.5m$.
With the size of the r-WINE array held constant, $d_{IRS}$ determines the number of installed IRS units, which equals the number of simultaneously deployable r-WINE links. The maximum number of r-WINE links is configured using bandwidth and channel propagation parameters from the link budget analysis in Section \ref{sec:link_budget}.
The resulting signal beams are ray-traced with 3dB beamwidth consistent with the antenna gains \cite{Rikkinen2020}. WINE-cellar tracks beam overlaps and intersections, obtaining the proportion of r-WINE links subject to the mutual interference. While a dense r-WINE array enhances the IRS panel availability, wireless signals transmitted in a compact area are susceptible to the interference. Conversely, a sparse r-WINE array reduces the chance of beam collisions but limits the number of IRS units.

Ruling out wireless signals impaired by beam collisions, the average throughput of the remaining r-WINE links is obtained from link budget analysis results in Section \ref{sec:link_budget}. Extreme values of $d_{IRS}$ result in either excessive interference or few available IRS units, indicating the presence of the optimal $d_{IRS}$ for a given HPC housing configuration. Such trends are observed with different heights of the machine room, denoted as $h$. Small values of $h$ yield short propagation distances to the ceiling, which is favorable for preserving the directionality and the energy of r-WINE signals. This also tightens the machine room space and increases the chance of beam collisions. Therefore, as $h$ becomes larger, the optimal $d_{IRS}$ becomes smaller since the interference is managed easily in a spacious environment. Notably, excessively high ceiling, e.g., $h=9.5m$, weakens the signal strength and lowers the overall utility.
Among the tested design cases, the configuration of $h=7.5m$ and $d_{IRS}=1.8m$ realizes the average r-WINE throughput of 300Gbps, showcasing the feasibility of the interference management and the integration of WINE in real-world HPC systems.

\subsection{WINE link coordination and computing performance enhancement}

The coordination of s-WINE, r-WINE, and wired links adapts the interconnect to a variety of HPC use cases, transforming the physical network structure into a logical structure that matches efficient workflows.
To observe how wireless functionalities enhance HPC networking and computing performances, a wired-only baseline interconnect is constructed with WINE-cellar. The baseline is structured via $k$-ary 3-cube torus, and different WINE-cellar instances are constructed to encompass scale-out configurations of $k \in \{4,6,8\}$.
Note that the framework extends to many other widely-adopted interconnect topologies. For example, wireless links can reinforce high-level bandwidths in fat-trees or provide remote inter-group connections of Dragonfly to reduce the transmission latency. Along the baseline interconnect, s-WINE and r-WINE links facilitate the direct communication between two racks separated apart by many hops in the wired structure. Wired links, borrowing the NDR InfiniBand specifications, require 0.6$\mu s$ single-hop latency.

\begin{figure}
    \centering
    \includegraphics[width=\linewidth]{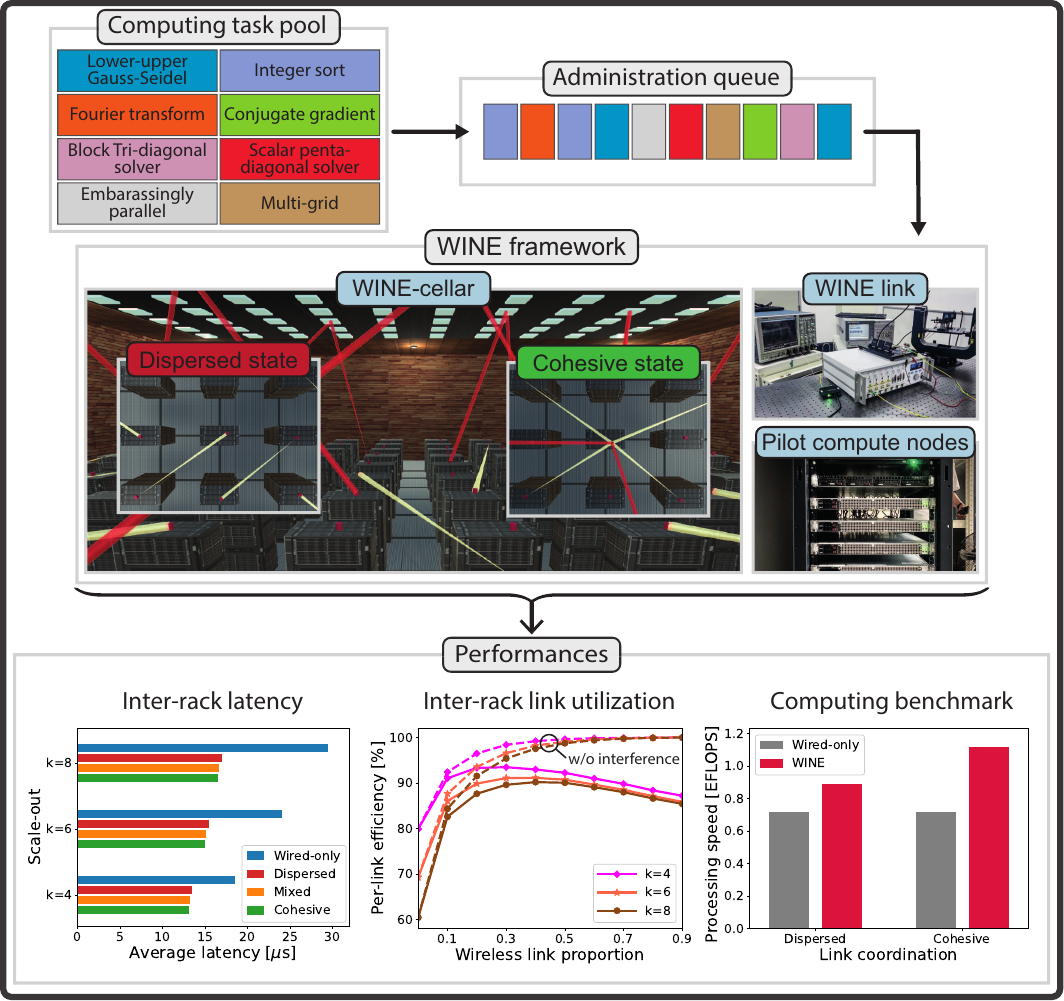}
    \caption{WINE link coordination and performances.}
    \label{fig:wine_coordination}
\end{figure}

Figure \ref{fig:wine_coordination} illustrates the link coordination and performances of WINE-integrated HPC system, along with wireless link testbed and pilot compute node platform. Widely-deployed HPC benchmark workloads \cite{npb} are serially administered to the WINE-cellar HPC prototype. These benchmarks address distinct patterns of inter-rack data sharing, providing a comprehensive appraisal of networking and computing utilities. In large-scale interconnects, such communication patterns significantly affect the communication latency \cite{random_topology_hpc}. For example, lower-upper Gauss-Seidel solver \cite{npb} involves data sharing mainly among neighboring racks. Coordinating wireless links among nearby racks enhances the local connectivity, resulting in a {\em dispersed} state that specifies in handling locally contained data sharing. In contrast, Fourier transform and integer sorting \cite{npb} entail global referencing of the information distributed across all racks. Major bottleneck in such global cooperation arises from the communication delay along the critical path of the network, or the tail latency \cite{tail_scale}. To overcome this, network science analysis \cite{barabasiTemporal} reveals that diameters of large graphs are shortened by high-degree {\em hub} nodes connected to many neighbors, providing short routes between arbitrary node pairs. This motivates the WINE link coordination to shape a {\em cohesive} state, where wireless connections are focused on hub racks.

The WINE link coordination assesses data traffic patterns of administered workloads and subsequently deploys s-WINE and r-WINE links for enhanced connectivity. The resulting interconnect structure is classified as dispersed, cohesive, or mixed, based on the distribution of neighboring connection counts. For all states, wireless links reduce the number of hops required for the inter-rack communication. This results in the decrease of average rack-to-rack latencies compared to the baseline. The latency reduction is particularly significant in the cohesive state, since hub racks provide short routes between remote non-neighboring racks. The near-half latency comparison with legacy topologies showcases the potential to overcome the inter-rack communication delay over large-scale HPCs.

Extensive job administration overburdens the interconnect, resulting in the network congestion that limits full utilization of the link bandwidth. Wireless links provide direct connections that bypass such congestion points, enhancing the average inter-rack data rate.
Since the supply for additional links trivially enhances the availability of communication routes, the average link utilization rate are assessed with varying proportions of wired and wireless links, with the total number of rack connections kept consistent.
When over 30\% of connections are configured wirelessly, increased mutual interference causes a slight decay in data rate. Despite this, the framework consistently upholds the link utilization efficiency above 80\%, surpassing that of the wired-only baseline even under moderate interference conditions. Advanced beamforming technologies provide small beamwidths to minimize beam collisions, further enhancing the throughput under dense wireless environments.
Notably, higher gains in the link utilization rates are obtained in more scaled-out systems, rendering the networking loads over future HPCs favorably manageable with the WINE framework.

The impact of WINE on the parallel processing speed is assessed using pilot HPC systems constructed by small-scale compute nodes equipped with wireless interfaces. These units are interconnected by cables in a topology that mirrors the 3-ary torus baseline of the digital twin. Under different scalings of $k\in\{2,3,4\}$, the average required number of hops for data sharing and computing benchmarks are obtained, both with the baseline connections only and with the utilization of wireless interfaces. For example, a 4-ary pilot system delivers the benchmark performance of 0.85GFLOPS with the baseline connections only. Upon the deployment of wireless interconnections to shape dispersed and cohesive states, the average number of hops for data sharing decreases from 3 to 2.87 and 2.67, improving the benchmark by 4.18\% and 11.13\%, respectively. Extrapolating these gains to the average processing speed of the top 5 systems on the TOP500 list \cite{top500}, WINE achieve up to 1,122PFLOPS upon the deployment in real-world top-tier HPC systems. This highlights the potential for surpassing the post-exascale barrier in HPC through wireless technologies.

\section{Conclusion and ongoing research opportunities}
\label{sec:conclusion}
This article presents a wireless technology-integrated interconnection framework, termed WINE, for the next-generation HPC systems. WINE deploys highly directional broadband signals and reflective environment controlling to enable reliable high-speed wireless data links among HPC racks. The coordination of s-WINE, r-WINE and wired links promotes the interconnect structure adaptability to diverse use cases, thereby enhancing key networking and computing performances. The feasibility analysis, conducted with the digital twin HPC testbed WINE-cellar, verifies the viability of practical deployment in modern state-of-the-art HPC designs. The consolidation of further studies and practical development sheds light on the following research challenges.

\begin{itemize}
    \item WINE link coordination strategy and protocol: WINE enables versatile controls over wired and wireless links of HPC interconnects. The collective administration of these link types yields countless interconnect structures, offering new ways to optimize processing workflows of diverse applications. Coordination protocols to find and implement resource-efficient job mappings, given the WINE-integrated interconnect structure, can exploit full potential of the framework. These coordination principles address practical issues in the link deployment, such as delays in adjusting transceivers to align with different directions of coherent signals.
    
    \item Dense wireless link management: The dense deployment of wireless links in compact HPC environments presents challenges in managing the signal obstruction and the mutual interference. Simulations have shown that advanced transceiver technologies, which realize narrow beam signals with high spatial resolution, can mitigate the issues. The flexibility in configuring wireless links with the minimal signal obstruction and interference can be further enhanced through new placement strategies for racks, transceivers, and IRS units. Deploying the racks on a curved-surface floor ensures additional line-of-sight connections for s-WINE links. Furthermore, enhanced reflective environment control strategies, possibly with the signal processing equipment for multi-step reflections, can be explored.

    \item HPC administration via digital twin: A digital twin for HPC system prototyping, called WINE-cellar, has been used to verify the feasibility of various design cases. Potential functionalities of WINE-cellar extend beyond prototyping and simulation, encompassing practical job and link administration of WINE-integrated HPC clusters. A pilot hardware platform, possibly through micro-ATX form-factor compute nodes and wireless transceivers synchronized with WINE-cellar configurations, can be built for the development of such administrative features. Such a setup allows for comprehensive testing and refinement of both software and hardware components in the realistic HPC environment.
\end{itemize}

\bibliographystyle{IEEEtran}
\bibliography{references}

\vskip -2\baselineskip plus -1fil

\begin{IEEEbiographynophoto}
{Hong Ki Kim} %(istackcheese@korea.ac.kr)
received the B.S. degree from the Korea University in 2019. He is currently with the School of Electrical Engineering, Korea University, Seoul, Korea. His research interests include computing cluster interconnection, machine-type communication-based agent cooperation, digital twin platform, and their applications.
\end{IEEEbiographynophoto}

\vskip -2\baselineskip plus -1fil

\begin{IEEEbiographynophoto}
{Yong Hun Jang} %(disclose@korea.ac.kr)
received the B.S. degree from the Seoul National University of Science and Technology in 2019. He is currently with the School of Electrical Engineering, Korea University, Seoul, South Korea. His research interests include distributed system optimization, computing cluster parallelization and job scheduling, intelligent reflecting surface for wireless communication and their applications.
\end{IEEEbiographynophoto}

\vskip -2\baselineskip plus -1fil

\begin{IEEEbiographynophoto}
{Hee Soo Kim} %(huolangheesoo@korea.ac.kr)
received the B.S. degree from the Korea University, Seoul, South Korea in  2017. He is currently pursuing the Ph.D. degree with the School of Electrical Engineering, Korea University. His research interests include HPC topology design through network science, and large-scale multi-slice networks.
\end{IEEEbiographynophoto}

\vskip -2\baselineskip plus -1fil

\begin{IEEEbiographynophoto}
{Won Young Kang} %(dogs0667@korea.ac.kr)
received the B.S. degree from the Hannam University in 2022. He is currently with the School of Electrical Engineering, Korea University, Seoul, South Korea. His research interests include wireless systems, millimeter-wave communications and signal processing for communications.
\end{IEEEbiographynophoto}

\vskip -2\baselineskip plus -1fil

\begin{IEEEbiographynophoto}
{Young-Chai Ko} %(koyc@korea.ac.kr)
received the B.S. degree in electrical and telecommunication engineering from Hanyang University, Seoul, South Korea, and the M.S.E.E. and Ph.D. degrees in electrical engineering from the University of Minnesota, Minneapolis, MN, in 1999 and 2001, respectively.
He is currently with the School of Electrical Engineering, Korea University, as a Professor. His current research interests include the design and evaluations of multi-user cellular systems, MODEM architecture, millimeter-wave, and THz wireless systems.
\end{IEEEbiographynophoto}

\vskip -2\baselineskip plus -1fil

\begin{IEEEbiographynophoto}
{Sang Hyun Lee} (sanghyunlee@korea.ac.kr)
received the B.S. and  M.S. degrees from the Korea Advanced Institute of Science and Technology in 1999 and 2001, respectively, and the Ph.D. degree from the University of Texas at Austin in 2011. He is currently with the School of Electrical Engineering, Korea University, Seoul, Korea. His research interests include communications, designs, and optimizations on wireless applications in high-performance computing.
\end{IEEEbiographynophoto}

\end{document}